\documentclass[useAMS,usenatbib,letterpaper]{mn2e}
\usepackage{graphicx}

\newcommand{\Ek}[1]{E_{{\rm k}, #1}}

\title[The KH instability in weakly ionised plasmas II]{The
	Kelvin-Helmholtz instability in weakly ionised plasmas II: 
		multifluid effects in molecular clouds.}
\author[A. C. Jones and T. P. Downes]{A.C. Jones$^{1,3}$ and 
	T. P. Downes$^{1,2,3}$\thanks{E-mail: turlough.downes@dcu.ie (TPD)}\\
$^{1}$School of Mathematical Sciences, Dublin City University,
	      Glasnevin, Dublin 9, Ireland\\
$^{2}$School of Cosmic Physics, Dublin Institute for Advanced
              Studies, 31 Fitzwilliam Place, Dublin 2, Ireland \\
$^{3}$National Centre for Plasma Science and Technology, Dublin
         City University, Glasnevin, Dublin 9, Ireland}

\begin{document}

\date{Accepted October 28, 2011. Received 18 October, 2011; in original
	form September 8, 2011}

\pagerange{\pageref{firstpage}--\pageref{lastpage}} \pubyear{----}

\maketitle

\label{firstpage}

\begin{abstract}

We present a study of the Kelvin-Helmholtz instability in a weakly ionised, multifluid MHD plasma with
parameters matching those of a typical molecular cloud.  The instability is capable of transforming 
well-ordered flows into disordered flows. As a result, 
it may be able to convert the energy found in, for example, bowshocks from stellar jets into the turbulent 
energy found in molecular clouds. As these clouds are weakly ionised, the ideal magnetohydrodynamic 
approximation does not apply at scales of around a tenth of a parsec or less. This paper extends the work of 
\citet{paper1} on the evolution of the Kelvin-Helmholtz instability in the presence of multifluid 
magnetohydrodynamic effects. These effects of ambipolar diffusion and the Hall effect are here studied 
together under physical parameters applicable to molecular clouds.  We restrict our attention to the 
case of a single shear layer with a transonic, but super-Alf\'enic, velocity jump and the computational
domain is chosen to match the wavelength of the linearly fastest growing mode of the instability.

We find that while the introduction of multifluid effects does not affect the linear growth rates of the instability, the non-linear behaviour undergoes considerable change. The magnetic field is decoupled from the bulk flow as a result of the ambipolar diffusion, which leads to a significant difference in the evolution of the field. The Hall effect would be expected to lead to a noticeable re-orientation of the magnetic field lines perpendicular to the plane. However, the results reveal that the combination with ambipolar diffusion leads to a surprisingly effective suppression of this effect.
\end{abstract}

\begin{keywords}
mhd -- instabilities -- ISM:clouds -- ISM:kinematics and dynamics
\end{keywords}

\section{Introduction}

The Kelvin-Helmholtz (KH) instability can occur anywhere that has a
velocity shear and, as a result, is an important instability
in almost any system involving fluids. The instability has been studied in a variety of astrophysical systems, 
from solar winds \citep{amerstorfer07, bettarini06, hasegawa04} and pulsar 
winds \citep{bucciantini06} to thermal flares \citep{venter06}. Due to its 
ability to drive mixing and turbulence, the KH instability has been considered 
relevant in protoplanetary disks \citep{johansen06, gomez05}, accretion disks 
and magnetospheres \citep{li04}, and other jets and outflows \citep{baty06}.

One environment in which the presence of turbulent energy is of particular interest is molecular clouds.  One possible source of this turbulent energy is the interaction of protostellar jets with the surrounding cloud, as first proposed by \citet{norman80}. The transfer of momentum from a jet to the cloud is most likely to occur through
so-called ``prompt entrainment'' \citep[e.g.][]{dyson84}. As the supersonic protostellar jet propagates into the surrounding cloud, it forms a bow
shock. This shock accelerates molecular cloud material, imparting momentum to it.  It is worth noting that the momentum imparted is relatively 
well ordered: a further process must occur to convert this momentum into turbulent motions.

Early studies of the KH instability in protostellar jets
were carried out using perturbative linear analysis. These initial
studies \citep[as reviewed by][]{birkinshaw91} examine the fluid
equations under set-ups of various combinations of magnetic field and
shear layers, etc. By approaching the problem using a simple
mathematical treatment, the effect of each physical parameter can be
followed closely. \citet{hardee97num}, for example, investigate the KH
instability in jets by solving the dispersion relations for KH modes
over a wide range of perturbation frequencies. However, these studies
are limited to the linear regime of the instability, while the evolution of 
stellar jets is very much governed by nonlinear phenomena \citep{bodo94}. Subsequent studies have followed the growth of the KH instability in jets using time-dependent numerical simulations. \citet{stone97} and \citet{downes98}, to name just a few, impose linear perturbations onto an initially stable set-up to observe the behaviour of the instability into the non-linear regime and the effect it can have in stellar jets.


While many authors have investigated the role of the KH instability in general
magnetised and unmagnetised astrophysical flows \citep[e.g.][]{frank96, mala96, 
hardee97num, downes98, keppens99}, these studies have investigated the 
KH instability in the context of either hydrodynamics or ideal magnetohydrodynamics (MHD).  
These assumptions are, however, not always valid, particularly in weakly ionised systems. For example, in molecular clouds we
know that non-ideal effects are important at length
scales below about 0.2\,pc \citep[e.g.][]{ois06, downes09} and hence it
is of interest to explore the KH instability in the context of either
non-ideal MHD or, preferably, fully multifluid MHD.  

In recent years the emphasis of KH studies has turned to including non-ideal
effects.  \citet{keppens99} studied both the linear growth and subsequent 
nonlinear saturation of the KH instability using resistive MHD numerical 
simulations. The inclusion of diffusion allowed for magnetic reconnection
and non-ideal effects were observed through tearing instabilities and  
the formation of magnetic islands. The case in support of using numerical diffusion in order to simulate non-ideal MHD effects was argued the following year by \citet{jeong00}, as analogous to the similar practice used to simulate non-ideal hydrodynamic flows of high Reynolds number. 

 \citet{birk02} examined the case of a partially 
ionised dusty plasma, using a multifluid approach in which collisions could be 
included or ignored.  They found that collisions between the neutral fluid and 
dust particles could lead to the stabilisation of KH modes of particular 
wavelengths. The unstable modes led to a significant local amplification of 
the magnetic field strength through the formation of vortices and current 
sheets. In the nonlinear regime they observed the magnetic flux being 
redistributed by magnetic reconnection. It was suggested that this could be 
applicable to dense molecular clouds and have important implications for the 
magnetic flux loss problem \citep{umebayashi90}.

A comprehensive study was carried out by \citet{wiechen06} which
demonstrated the effect of dealing with the plasma using a multifluid 
scheme. This study focused on the effect of varying the properties of 
the dust grains. The results of the simulations led to the conclusions that 
more massive dust grains have a stabilising effect on the system while higher 
charged numbers have a destabilising effect. It was found that there is no 
significant dependence on the charge polarity of the dust.

\citet{palotti08} also carried out a series of simulations using
resistive MHD. They found that, following its initial growth, the KH
instability decays at a rate that decreases with decreasing plasma
resistivity, at least within the range of resistivities accessible to
their simulations. They also found that magnetisation increases the
efficiency of momentum transport, and that the transport increases with
decreasing resistivity. 

In \citet[henceforth Paper I]{paper1} we examine the 
behaviour of the KH instability in the presence of multifluid effects. 
We found that, while the linear growth rates of the instability are unaffected 
by multifluid effects, the non-linear behaviour was remarkably different. The inclusion of 
ambipolar diffusion leads to the removal of large quantities of magnetic energy while the Hall 
effect, if strong enough, proved capable of introducing a dynamo effect. This leads to continuing 
strong growth of the magnetic field well into the non-linear regime and
a lack of true saturation of the instability.

In this paper we perform a numerical simulation of the complete evolution
of the KH instability in a weakly ionised, multifluid plasma as found in molecular clouds. 
We address the case of a shear layer with a transonic velocity difference.
We choose a magnetic field strength typical of dense molecular clouds and this yields a
velocity difference across the shear layer which is highly super-Alfv\'enic (see section
\ref{sec:num-setup}).  It is worth noting here that the precise value of the Alfv\'en number is known to influence 
the evolution of the instability \citep[e.g.][]{jones97, baty03}.  The work here can be most
directly compared to Case 4 of \citet{jones97}, although it should be borne in mind that our
boundary conditions are slightly different.  The
molecular cloud material is simulated by four individual fluids: a neutral fluid, an electron fluid, a positively charged metal ion fluid and a fluid of large, negatively charged dust grains.  The non-ideal effects of ambipolar
diffusion and the Hall effect are included in the simulation, and their effect on the linear development, saturation and subsequent
behaviour of the instability is analysed in detail.  

The aim of this work is to investigate the growth and saturation of the KH 
instability under the influence of the multifluid effects found in molecular 
clouds.  
The KH instability is of particular interest as a possible means of converting the ordered energy injected into the cloud by protostellar jets into the turbulent energy observed. In Paper I we ran simulations with parameters chosen to simulate very high, medium and
very low magnetic Reynolds number systems and with parameters chosen to ensure 
ambipolar-dominated flows and Hall-dominated flows in order to
develop a full understanding of the roles of each. In this paper, we investigate the combined effects of these two multifluid effects on the instability under parameters that describe the specific environment of molecular clouds and compare these to roles played by each on the development of the instability as studied in Paper I. In doing so, we determine the behaviour of the instability in a physical application which, in practice, should prove observable.

In section \ref{sec:num-setup} we outline the multifluid equations used by the code for this analysis, the physical model
being simulated, and the computational parameters employed. In section \ref{sec:validation-numerics} we describe how the ability of the code to simulate the KH instability has been validated, in both the cases of ideal MHD and in the presence of multifluid effects. In section \ref{sec:results} we detail the 
results in both the linear and non-linear regimes, indicating where the non-ideal behaviour can be attributed to individual effects 
of ambipolar diffusion and the Hall effect, or the combination of the two.

\section{Numerical setup}
\label{sec:num-setup}

The simulations described in this work are performed using the {\sevensize HYDRA} code 
\citep{osd06, osd07} for multifluid magnetohydrodynamics in the weakly 
ionised regime. We further assume the flow is isothermal.  The assumption of 
weak ionisation allows us to ignore the inertia of the charged species and 
allows us to derive a (relatively) straightforward generalised Ohm's law.  The 
resulting system of equations, given below, incorporates finite parallel, Hall and
Pederson conductivity and, as such, is valid in molecular clouds. In these regions the viscous lengthscales are much smaller than those over which non-ideal effects are important. This leads to high Prandtl numbers and plasma flows in these regions can be considered to be effectively inviscid. 

Although protostellar jets in molecular clouds tend to be hypersonic \citep{ray88}, this study would be
appropriate to the KH instability which might arise due to the shear between the transonic flow of the 
swept-up material along the edge of the bowshock and the ambient medium, or to the shear along the contact
discontinuity just behind the bowshock. It has been found that the KH instability plays a greater role in the transonic, rather than the supersonic, regime \citep{miura90,miura92,koba08,frank96}.
Thus the low Mach number flows, taken in concert with the 
isothermal assumption, means that features in the flow such as shocks are 
unlikely to create regions of high ionisation.

In this work we compare the results of three simulations: the first is a
full, multifluid MHD simulation of the development of the KH instability
in a plasma with molecular cloud properties, the second is of the
development of the KH instability in the ideal MHD limit, and the third is a simulation of the development of
the instability in a hydrodynamic system.

\subsection{Multifluid equations}
\label{subsec:multifluid-eqns}

The code {\sevensize HYDRA} solves the following equations for a system of $N$
fluids. The simulations described in this paper consist of $N=4$ fluids,
	indexed by $i=0$ for the neutral fluid and $i=1, 2, 3$ for the electron, ion, and dust grain fluids respectively. The equations to be solved are

\begin{equation}
\frac{\partial \rho_i}{\partial t} + \mathbf{\nabla} \cdot (\rho_i \mathbf{q}_i) = 0 \hspace{0.5cm} (0 \le i \le N-1),
\end{equation}

\begin{equation}
\frac{\partial \rho_0 \mathbf{q}_0}{\partial t} + \mathbf{\nabla} \cdot (\rho\mathbf{q}_0\mathbf{q}_0 + c_{\rm s}^2\rho\mathbf{I}) = \mathbf{J} \times \mathbf{B},
\end{equation}

\begin{equation}
\frac{\partial \mathbf{B}}{\partial t} + \mathbf{\nabla} \cdot (\mathbf{q}_0 \mathbf{B} - \mathbf{B} \mathbf{q}_0) = - \mathbf{\nabla} \times \mathbf{E}^\prime,
\end{equation}

\begin{equation}
\alpha_i \rho_i (\mathbf{E} + \mathbf{q}_i \times \mathbf{B}) + \rho_i\rho_0K_{i0}(\mathbf{q}_0 - \mathbf{q}_i) = 0  \hspace{0.2cm} (1 \le i \le N-1),
\end{equation}

\begin{equation}
\mathbf{\nabla} \cdot \mathbf{B} = 0,
\end{equation}

\begin{equation}
\mathbf{\nabla} \times \mathbf{B} = \mathbf{J},
\end{equation}

\begin{equation}
\sum_{i=1}^{N-1}\alpha_i\rho_i = 0,
\end{equation}

\noindent where $\rho_i$, $\mathbf{q}_i$, $\mathbf{B}$, and $\mathbf{J}$
are the mass density and velocity of fluid $i$, the magnetic field and current density, respectively. $c_{\rm s}$ denotes the sound speed, and $\alpha_i$ and $K_{i0}$  are the charge-to-mass ratios and the collision coefficients between the charged species $i$ and the neutral fluid, respectively.

These equations lead to an expression for the electric field in the frame of the fluid, $\mathbf{E}^\prime$, given by the generalised Ohm's Law
\begin{equation}
\mathbf{E}^\prime = \mathbf{E}_{\rm O} + \mathbf{E}_{\rm H} + \mathbf{E}_{\rm A},
\end{equation}
where the components of the field are given by
\begin{equation}
\mathbf{E}_{\rm O} = (\mathbf{J} \cdot \mathbf{a}_{\rm O})\mathbf{a}_{\rm O},
\end{equation}

\begin{equation}
\mathbf{E}_{\rm H} = \mathbf{J} \times \mathbf{a}_{\rm H},
\end{equation}

\begin{equation}
\mathbf{E}_{\rm A} = -(\mathbf{J} \times \mathbf{a}_{\rm H}) \times \mathbf{a}_{\rm H},
\end{equation}

\noindent using the definitions $\mathbf{a}_{\rm O} \equiv f_{\rm O}\mathbf{B}$, $\mathbf{a}_{\rm H} \equiv
f_{\rm H}\mathbf{B}$, $\mathbf{a}_{\rm A} \equiv f_{\rm A}\mathbf{B}$, where $f_{\rm O} \equiv \sqrt{r_{\rm
	O}}/B$, $f_{\rm H} \equiv r_{\rm H}/B$ and $f_{\rm A} \equiv \sqrt{r_{\rm A}}/B$. The resistivities given
	here are the Pederson, Hall and ambipolar resistivities, respectively, and are defined by

\begin{equation}
r_{\rm O} \equiv \frac{1}{\sigma_{\rm O}},
\end{equation}

\begin{equation}
r_{\rm H} \equiv \frac{\sigma_{\rm H}}{\sigma_{\rm H}^2 + \sigma_{\rm A}^2},
\end{equation}

\begin{equation}
r_{\rm A} \equiv \frac{\sigma_{\rm A}}{\sigma_{\rm H}^2 + \sigma_{\rm A}^2},
\end{equation}
where the conductivities are given by

\begin{equation}
\sigma_{\rm O} = \frac{1}{B} \sum_{i=1}^{N-1} \alpha_i\rho_i\beta_i,
\end{equation}

\begin{equation}
\sigma_{\rm H} = \frac{1}{B} \sum_{i=1}^{N-1} \frac{ \alpha_i\rho_i}{1+\beta_i^2},
\end{equation}

\begin{equation}
\sigma_{\rm A} = \frac{1}{B} \sum_{i=1}^{N-1} \frac{ \alpha_i\rho_i\beta_i}{1+\beta_i^2},
\end{equation}
where the Hall parameter $\beta_i$ for a charged species is given by 
\begin{equation}
\beta_i = \frac{\alpha_i B}{K_{i0}\rho_0} .
\end{equation}

To solve these equations numerically we use three different operators:
\begin{enumerate}
\item solve equations (1), (2), (3), including the restriction of
	equation (5) and for $i=0$, using a standard second order, finite volume 
	shock-capturing scheme.  Note that for this operator the resistivity terms 
	in equation (3) are not incorporated.  Equation (5) is incorporated using 
	the method of Dedner \citep{dedner}.
\item Incorporate the resistive effects in equation (3) using 
super-time-stepping to accelerate the ambipolar diffusion term and the
Hall Diffusion Scheme to deal with the Hall term.
\item Solve equations (4) for the charged species velocities and use
these to update equation (1) with $i=1,\ldots,N-1$.
\end{enumerate}
These operators are applied using Strang operator splitting in order to
maintain the second order accuracy of the overall scheme.  We refer the
reader to \cite{osd06, osd07} for a more detailed description.

\subsection{Physical parameters}
\label{subsec:physical-parameters}

This study aims to simulate the multifluid effects on the KH instability in the environment of a molecular cloud. In a jet-driven bowshock for example, both the swept-up material in the bowshock, as well the surrounding region, are understood to be made up of molecular cloud material. For this reason, the physical parameters observed in molecular clouds are applicable to this system.

\subsubsection{Properties of bulk fluid}
\label{sec:fluid-props1}

The molecular cloud material is modelled after the single-size grain
model in \citet{wardle99} \citep[see][for further details]{jones11}. The plasma is weakly ionised, and therefore consists primarily of a neutral particle fluid. The remainder of the plasma is made up of an electron fluid, a positively charged metal ion fluid, and a fluid of large, negatively charged dust grains. The average mass of the neutral particles is taken to be $m_0 = 2.33\,m_{\rm p}$, where $m_{\rm p}$ is the mass of a proton. This corresponds to a fluid of 90\% molecular hydrogen and 10\% atomic helium by number, which is representative of molecular clouds. The ion fluid represents an average of ions produced from a number of metal atoms, including Na, Mg, Al, Ca, Fe and Ni. These metals have sufficiently similar ionisation and recombination rates, and so can be modelled collectively as ions of a single positive charge $+e$ \citep{umebayashi90}. The molecular ions, of which HCO\textsuperscript{+} is the most numerous, are significantly less abundant than the metal ions, allowing us to neglect them. An average mass of $m_2 = 24\,m_{\rm p}$ is assigned to the particles of the ion fluid, approximately equal to that of a magnesium ion. In the single-size grain model, the dust grains are characterised by a grain radius $r=0.1\,\mu \mbox{m}$ and a total grain mass that is 1\% of the total neutral mass. 

The simulations are parametrised by several properties that describe the overall system. The sound speed in the plasma is calculated from 
\begin{equation} c_{\rm s} \equiv \sqrt{\frac{\gamma k_{\rm B} T}{m_0}} ,\end{equation}
where $k_{\rm B} = 1.38 \times 10^{-16} \, \mbox{erg} \, \mbox{K}^{-1}$ is the Boltzmann constant, the adiabatic index $\gamma$ is equal to unity for an isothermal system, and the constant temperature is taken to be $T = 30 \, \mbox{K}$, which is consistent with molecular clouds \citep{wardle99}. This results in a sound speed of $3.26 \times 10^4 \, \mbox{cm} \, \mbox{s}^{-1}$, or $c_{\rm s} = 0.326 \, \mbox{km} \, \mbox{s}^{-1}$.

The plasma flow in the simulations is required to be transonic in order to allow for reasonable growth of the instability. The sonic Mach number is set as:
\begin{equation} M_{\rm s} = \frac{V_0}{c_{\rm s}} = 1,\end{equation}
so that the relative velocity between the bowshock and ambient medium is equal to the sound speed. In the reference frame of the instability, the velocity of the bulk fluid in the plasma is given as $\pm \frac{V_0}{2} = \pm 0.188 \, \mbox{km} \, \mbox{s}^{-1}$. 

The strength of magnetic fields in molecular clouds can vary greatly, but is generally observed as a few tens of microGauss \citep{zweibel95}. The magnitude of the magnetic field in this study is chosen as $|B_0| = 22.8 \, \mu \mbox{G}$ so that the Alfv\'{e}n velocity is:
\begin{equation} v_{\rm A} = \sqrt{\frac{B_0^2}{4 \pi \rho}} = 0.0326 \, \mbox{km} \, \mbox{s}^{-1}.\end{equation}
This gives an Alfv\'{e}n Mach number of 
\begin{equation} M_{\rm A} = \frac{V_0}{v_{\rm A}} = 10,\end{equation}
so that the plasma is super-Alfv\'{e}nic. It should be noted that there exists a wide range of fluid
velocities and magnetic field strengths in molecular clouds, in which the average thermal and magnetic
energies are approximately in equipartition. However, for this study, we have confined ourselves to regions of
transonic and super-Alfv\'{e}nic flows as the KH instability in a sub-Alfv\'{e}nic flow would be 
stabilised by the magnetic field \citep{chand61}. This study is also applicable to regions with 
stronger magnetic fields that are not parallel to the plane of plasma flow, but in which the strength 
of the magnetic field as projected onto the plane of the plasma flow is super-Alfv\'{e}nic 
\citep{jones97}.  Our choice of a transonic, super-Alfv\'enic flow is similar, though not
identical, to Case 4 of \citet{jones97} and is thus not in the ``very weak'' field regime, but
rather the weak field regime.   In this regime the magnetic field can be initially wound up by the 
growth of the instability and then becomes subject to reconnection due to the ensuing field
reversals.  The final state of the system in this case, at least in ideal MHD, is that of a somewhat disordered,
widened shear layer which is no longer subject to the KH instability on wavelengths permitted by
the computational set-up.

\subsubsection{Fluid masses and densities}
\label{sec:fluid-props2}

A typical number density of particles in a molecular cloud is $n \approx n_0 = 10^6 \, \mbox{cm}^{-3}$ \citep{draine83}. The number density of the remaining fluids can then be derived from figure 2 in \citet{umebayashi90}. The number density of the grain fluid is calculated to be $n_3 \approx 1.85 \times 10^{-14}\,n_0$. The number density of the ion and electron fluids, according to \citet{umebayashi90}, are seen to be approximately $10^{-8} n_0$. However, these parameters are for a molecular cloud with a magnetic field of magnitude 1\,mG. In order to obtain similar conductivities as \citet{wardle99} within the cloud but for a weaker magnetic field, the number density of the ion fluid is taken to be $n_2 \approx 1.02 \times 10^{-11}\,n_0$. The number density of the electron fluid is not set explicitly at this point, it is instead defined by the requirement that charge neutrality holds at all times. It will be seen that the number and mass density of the electron fluid is small, as expected. 

The mass densities of the three dominant fluids can then be calculated. It is found, for the neutral fluid, that 
\begin{equation} \rho_0 \equiv m_0 n_0 = 3.89 \times 10^{-18} \, \mbox{g} \, \mbox{cm}^{-3}.\end{equation}
For the ion fluid, 
\begin{equation} \rho_2 \equiv m_2 n_2 = 4.08 \times 10^{-28} \, \mbox{g} \, \mbox{cm}^{-3}.\end{equation}
The total mass of the dust grains in molecular clouds is taken to be equivalent to 1\% of the total neutral mass \citep{wardle99}. This equates to $\rho_3 = 0.01\,\rho_0  = 3.89 \times 10^{-20} \, \mbox{g} \, \mbox{cm}^{-3}$. Equivalently, it can be written that 
\begin{equation} m_3 = 0.01 \frac{2.33 m_{\rm p}}{1.85 \times 10^{-14}} = 1.26 \times 10^{12}\,m_{\rm p}.\end{equation}

The charge-to-mass ratios for the three charged fluids can be calculated from their charges and masses, as given above, and converted into cgs units using $1\,\mbox{C} \approx 3 \times 10^9\,\mbox{statC}$. All charged fluids are taken to be singly-charged, and their charge-to-mass ratios are thus found to be 
\begin{equation}  \alpha_2 \equiv \frac{+1\,e}{m_2} = \frac{1.6 \times 10^{-19}\,\mbox{C}}{24 m_{\rm p}}  
= 1.2 \times 10^{13} \,\mbox{statC}\,\mbox{g}^{-1},\end{equation}

\begin{equation} \alpha_3 \equiv \frac{-1\,e}{m_3} 
= -2.28 \times 10^{2}\,\mbox{statC}\,\mbox{g}^{-1} , \end{equation}
and 
\begin{equation} \alpha_1 \equiv \frac{-1\,e}{m_1} = \frac{-1\,e}{m_{\rm e}} 
= - 5.27 \times 10^{17} \,\mbox{statC}\,\mbox{g}^{-1} . \end{equation}
From these values, the necessary mass density for the electron fluid can be calculated from 
\begin{equation} \alpha_1\rho_1 + \alpha_2\rho_2 + \alpha_3\rho_3  = 0\end{equation}
to give $\rho_1 = 9.25 \times 10^{-33} \,\mbox{g}\,\mbox{cm}^{-3}$. This is much smaller than the mass density of the other three fluids, as expected.

\subsubsection{Properties of charged fluids}
\label{sec:fluid-props3}

The multifluid effects are implemented in the system through collisions between the various fluids. Due to the assumption of weak ionisation, only collisions between the neutral fluid and each charged fluid are included. Rate coefficients for momentum transfer for each fluid interaction are given in \citet{wardle99}. For collisions between the neutral and dust grain fluids, the collision rate is
\begin{equation} K_{\rm dust, \rm n} \equiv \frac{<\sigma v>_3}{m_3 + m_0} ,\end{equation}
where 
\begin{equation} <\sigma v>_3 = \pi r^2 \left( \frac{128 k T}{9 \pi m_0} \right)^{1/2} = 2.18 \times 10^{-5} \,\mbox{cm}^3 \mbox{s}^{-1} . \end{equation}
The collision rate can then be calculated as 
\begin{eqnarray} K_{3,0} &\equiv& \frac{<\sigma v>_3}{m_3 + m_0} = \frac{2.18 \times 10^{-5} \,\mbox{cm}^3 \mbox{s}^{-1} }{1.26 \times 10^{12} m_{\rm p} + 2.33 m_{\rm p}} \nonumber \\
&=& 1.0 \times 10^7 \,\mbox{cm}^3 \mbox{g}^{-1}\mbox{s}^{-1} . \end{eqnarray}
Similarly, for the ion and electron fluids, it can be found that 
\begin{eqnarray} K_{2,0} &\equiv& \frac{<\sigma v>_2}{m_2 + m_0} = \frac{1.6 \times 10^{-19} \,\mbox{cm}^3 \mbox{s}^{-1} }{24 m_{\rm p} + 2.33 m_{\rm p}} \nonumber \\
& =& 3.64 \times 10^{13} \,\mbox{cm}^3 \mbox{g}^{-1}\mbox{s}^{-1} \end{eqnarray}
and finally, following \citet{wardle99},
\begin{eqnarray} K_{1,0} &\equiv& \frac{<\sigma v>_1}{m_1 + m_0} = \frac{1 \times 10^{-15} cm^2 \left( \frac{128 k T}{9 \pi m_{\rm e}} \right)^{1/2} }{m_{\rm e} + 2.33 m_{\rm p}} \nonumber \\
&=& 1.16 \times 10^{15} \,\mbox{cm}^3 \mbox{g}^{-1}\mbox{s}^{-1} . \end{eqnarray}

Having determined the primary parameters of the fluids, the secondary parameters can be calculated. The Hall
parameter is a measure of how well tied a particle is to the magnetic field and is defined by $\beta_i \equiv \frac{\alpha_i (\frac{B}{c})}{K_{i,0} \rho_0}$. Using the quantities above, the Hall parameters for each fluid can be calculated. 

\begin{equation} \beta_2 = \frac{(1.2 \times 10^{13})(\frac{2.28 \times 10^{-5}}{3 \times 10^{10}})}{(3.64
		\times 10^{13})(3.89 \times 10^{-18})} = 6.43 \times 10^{1} \label{eqn:ion_hall} \end{equation}
\begin{equation} \beta_3 = \frac{(-2.28 \times 10^{2})(\frac{2.28 \times 10^{-5}}{3 \times 10^{10}})}{(1.0
		\times 10^7)(3.89 \times 10^{-18})} = -4.3 \times 10^{-3} \label{eqn:dust_hall} \end{equation}
\begin{equation} \beta_1 = \frac{(-5.27 \times 10^{17})(\frac{2.28 \times 10^{-5}}{3 \times 10^{10}})}{(1.16
		\times 10^{15})(3.89 \times 10^{-18})} = -8.92 \times 10^4 \label{eqn:electron_hall} \end{equation}

The final calculations are to solve for the conductivities. From the
definitions given in Sect. \ref{subsec:multifluid-eqns}, the Pederson, Hall and ambipolar resistivities can be calculated as:
\begin{equation} \sigma_O \equiv \frac{1}{B} \sum_{i=1}^{N-1} \alpha_i \rho_i \beta_i = 413.6 \, \mbox{s}^{-1} \end{equation}
\begin{equation} \sigma_H \equiv \frac{1}{B} \sum_{i=1}^{N-1} \frac{\alpha_i \rho_i}{1 + \beta_i^2} = 0.01 \, \mbox{s}^{-1} \end{equation}
\begin{equation} \sigma_A \equiv \frac{1}{B} \sum_{i=1}^{N-1} \frac{\alpha_i \rho_i \beta_i}{1 + \beta_i^2} = 0.1 \, \mbox{s}^{-1} . \end{equation}
It can be seen that these values are equal to those calculated by \citet{wardle99} for a weakly ionised molecular cloud.

It should be noted that this system consists of three charged fluids, while our multifluid KH study in Paper I
consisted of only two. In the previous study, the ion fluid was responsible for both the ambipolar and the
Hall conductivity. With three charged fluids in this study, the ion fluid is now responsible only for the
ambipolar conductivity, while the dust grain fluid provides the largest contribution to the Hall conductivity.
This results in a negative value for the the Hall conductivity, as the dust grains are negatively charged. The
only consequence of this is that the re-orientation experienced by the magnetic field as a result of the Hall effect is now in the opposite $z$-direction to before.

\subsection{Normalised computational parameters}
\label{subsec:computational-parameters}

The physical parameters for molecular clouds, outlined above, are
transformed into dimensionless units before being passed to the numerical code. For this purpose, practical characteristic length, time and mass scales must be determined. 

The lengthscale of the system is first chosen. The protostellar jet is understood to be approximately a parsec or two in length \citep{ray87}. It is chosen to focus on a fraction of this length, opting to simulate a region along the edge of the bowshock of length 0.2\,pc. It has also been found that multifluid effects play an important role at this lengthscale \citep{downes09}. The lengthscale is thus set to
\begin{equation} [L] = 0.2\,\mbox{pc} = 6.17 \times 10^{17}\,\mbox{cm}. \end{equation}
The timescale is chosen such that the sound speed is of order unity. The sound speed has been calculated as $c_{\rm s} = 3.26 \times 10^4 \,\mbox{cm} \,\mbox{s}^{-1}$. This implies a time scale 
\begin{equation}[T] = 1.9 \times 10^{13} \,\mbox{s}. \end{equation}
Using these units, the velocity of the plasma flow is now 
\begin{equation} \mathbf{v} = \pm \frac{V_0}{2} \hat{y} = \pm \frac{1}{2} \hat{y} . \end{equation}

Finally, a mass scale is chosen. This is set so that the mass density of the plasma is of order unity. The mass density has been calculated as $\rho_0 = 3.89 \times 10^{-18} \, \mbox{g} \, \mbox{cm}^{-3}$. Using the length scale chosen above, this gives a characteristic mass scale 
\begin{equation} [M] = 9.15 \times 10^{35} \,\mbox{g} . \end{equation}

The magnitude of the magnetic field, $B_0 = 22.8\, \mu \mbox{G}$, can also be transformed into a dimensionless quantity, by
\begin{equation} | \mathbf{B} | = \frac{B_0}{\sqrt{4 \pi}} \times \frac{[T] [L]^{1/2}}{[M]^{1/2}} = 0.1 \end{equation}

From this calculation onwards, the units of $\mathbf{B}$, $\mathbf{E}$ and $\mathbf{J}$ are written such that the factors of $4 \pi$ and the speed of light no longer appear.

It is possible now to transform the conductivities, and the resulting resistivities, into dimensionless units. The following values are found:

\begin{equation} r_O = \frac{1}{\sigma_O} = 6.2 \times 10^{-9} \end{equation}
\begin{equation} r_H = \frac{\sigma_H}{\sigma_H^2 + \sigma_A^2} = -0.00352 \end{equation}
\begin{equation} r_A = \frac{\sigma_A}{\sigma_H^2 + \sigma_A^2} = 0.0352 \end{equation}
These values can be seen to correspond to those implemented in Paper I. The amount of ambipolar resistivity is equivalent to the simulation with high ambipolar resistivity in Paper I (ambi-high-hr) with magnetic Reynolds number $Re_{\rm m} = 28.4$, while the amount of Hall resistivity is equivalent to the simulation with moderate Hall resistivity in Paper I (hall-med-hr) with magnetic Reynolds number $Re_{\rm m} = 284$.

\subsection{Grid initialisation and properties}
\label{subsec:computational-domain}

As in Paper I, these simulations are carried out on a 2.5\,D slab grid in the $xy$-plane. 
The initial set-up used was that of two plasmas flowing anti-parallel
side-by-side on a grid of size $x = [0, 32L]$ and $y = [0, L]$. The
plasma velocities are given by $+ \frac{V_0}{2}$ and $-\frac{V_0}{2}$ in
the $y$-direction. The unit of time used is the sound crossing time, $t_{\rm s} \equiv \frac{L}{c_{\rm s}}$, the time taken for a signal travelling at the speed of sound to cross the grid in the $y$-direction. 

We use periodic boundary conditions at the high and low $y$ boundaries.  
Since we wish to study not only the initial growth phase of the instability, 
but also its subsequent non-linear behaviour we must ensure that waves interacting
with the high and low $x$ boundaries do not reflect back into the domain
to influence the dynamics.  Previous test simulations for various
parameters have shown that a large width of $32L$ is necessary to
ensure this.  We use gradient zero boundary conditions at the high and
low $x$ and $z$ boundaries.

The grid is chosen therefore to consist of $6400 \times 200 \times 1$ cells, in the $x$, $y$,
and $z$ directions respectively. This resolution was chosen on the basis
that it reproduces the initial linear growth of the ideal MHD system in 
\cite{keppens99}.  Resolution studies were performed to confirm the
resolution as being appropriate for multifluid MHD as well (see Sect.\ \ref{sec:validation-resolution}).

The plasma velocity field is initiated with a tangential shear layer of width $2a$ at the
interface at {\bfseries $x = 16L$}. This velocity profile is described by
{\bfseries \begin{equation}
\mathbf{v}_0 =  \frac{V_0}{2} \tanh \left( \frac{x - 16L}{a} \right)
	\hat{\mathbf{y}} .
\end{equation}}
\noindent The width of the shear layer is chosen to be $\frac{a}{L} =
0.05$, or approximately 20 grid zones.  The magnetic field is initially
set to be uniform and aligned with the plasma flow.  

The initial background for all four fluids in the system is now an exact equilibrium. The initial velocity field, $V_0$ is then augmented with a perturbation
given by
\begin{equation}
\delta v_x = \delta V_0 \sin( - k_y y) \exp \left( - \frac{(x -
			16L)^2}{\sigma^2} \right).
\end{equation}
where $\delta V_0$ is set to $10^{-4}\,V_0$.  The wavelength of the
perturbation is set equal to the characteristic length scale, $\lambda
= \frac{2 \pi}{k_y} = L$, so that a single wavelength fits exactly into
the computational domain. This maximises the possibility of resolving
structures that are small relative to the initial perturbed wavelength,
\citep{frank96}. The perturbation attenuation scale is chosen so that it is 
larger than the shear layer, but small enough so that the instability can be 
assumed to interact only minimally with the $x$-boundaries 
\citep[see][]{palotti08}, and is set using $\frac{\sigma}{L} = 0.2$ 
\citep[see][]{keppens99, palotti08}. Finally, the wavenumber $k_y$ is chosen to be $2 \pi$ in order to
maximise the growth rate of the instability \citep{keppens99}. 

We note that our choice of periodic boundary conditions on the $y$ boundaries limits the behaviour of the 
system to some extent: wavelengths longer than the grid domain can not grow and coalescence of large-scale vortices 
will not occur \citep[e.g.][]{baty03}.  Furthermore, our choice of gradient zero boundary conditions in the
$x$ direction implies that we are studying only the surface modes of the instability.  This latter point, however, 
is not a major restriction as the body modes will not be important in, for example, the development of the KH 
instability in the bowshocks created by young stellar outflows.

\section{Validation of numerical approach}
\label{sec:validation-numerics}

\subsection{Validation of instability growth in ideal MHD}
\label{subsec:validation-KH}

The set-up described above has been shown in Paper I to be valid for producing the KH
instability under the ideal MHD approximation through comparisons with previously published literature. Both
hydrodynamic and ideal MHD simulations were run using the {\sevensize HYDRA} code. Comparisons between the
linear growth of the simulated hydrodynamic instability and the growth rate calculated analytically for the
same wavenumber by \citet{miura82} showed exceptional agreement. In the non-linear regime, the maximum reached
by the magnetic energy in the ideal MHD simulation matches that of \citet{mala96} to within 10\%.  This allows
us to be confident of the behaviour of {\sevensize HYDRA} in simulating the KH instability.

\subsection{Validation of instability growth in multifluid MHD}
\label{sec:validation-resolution}

The inclusion of multifluid effects introduces new length scales into
the system. These include the diffusion length scales of the magnetic
field due to ambipolar resistivity, and the rather computationally challenging 
whistler waves arising from the Hall effect. The Hall term is 
handled by {\sevensize HYDRA} using the explicit Hall
Diffusion Scheme (HDS) \citep{osd06, osd07}. Although the code naturally
does not resolve waves of vanishing wavelength, resolution studies were
performed in Paper I of both ambipolar and Hall-dominated flows and the 
results indicated that a resolution of $6400\times200\times1$ is sufficient 
to capture the initial growth and saturation of the instability.  
Subsequently, the dynamics are captured at least qualitatively. 

In order to ensure that the smallest-scale dispersive effects were in place when examining whether they are sufficiently resolved, the highest values for the ambipolar and Hall resistivity used in Paper I were implemented for these resolution studies. As indicated in Sect.\ \ref{subsec:computational-parameters}, this highest value of ambipolar resistivity in Paper I is equivalent to the value implemented in this paper (with magnetic Reynolds number $Re_{\rm m} = 28.4$), while the amount of Hall resistivity implemented in this study is equivalent to the simulation with only moderate Hall resistivity from Paper I (with magnetic Reynolds number $Re_{\rm m} = 284$).

We are, therefore, confident that the multifluid dynamics resulting from the non-ideal effects in these simulations are well 
resolved and that our conclusions as to the physical processes occurring are 
well-founded.

\section{Results}
\label{sec:results}

In order to observe the differences in the evolution of the KH instability with the inclusion of multifluid
effects, several aspects of the evolution are examined and compared to those from simulations carried out in
ideal MHD and in pure hydrodynamics \citep[see][for further details of the simulation results]{jones11}. 

The study of the growth of the instability is carried out through measuring the evolution of a
number of parameters with time. In particular, we
measure the transverse kinetic energy $$\Ek{x} \equiv \int \! \int 
\frac{1}{2} \rho v_x^2\,dx\,dy$$ and the magnetic energy $$E_b \equiv
\frac{1}{2} \int \! \int \left\{\left[ B_x^2 + B_y^2 + B_z^2
\right] - B_0^2\right\}\,dx\,dy$$ in the
system where $B_0$ is the magnitude of the magnetic field at $t=0$. 
Any growth of $\Ek{x}$ is due to the growth of the instability, as the entire plasma flow is initially in the $y$-direction, with only a very 
small perturbation in the $x$-direction,

\subsection{Linear regime}
\label{subsec:linear}

\begin{figure}
\centering
\includegraphics[width=8.4cm]{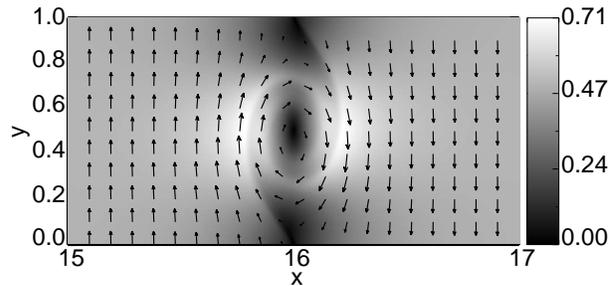}
  \caption{Plot of the magnitude and vector field of the bulk velocity field for the multifluid MHD case, at time of saturation of the instability, $t = 8 \, t_{\rm s}$.} 
 \label{weak_32wide_NI3_v_80.eps}
\end{figure}

The KH instability leads to an interaction between the two plasmas on either side of the initial interface. In
particular, in ideal MHD, the plasmas are seen to wind-up, leading to the ``Kelvin's cat's eye'' vortex. The
inclusion of multifluid effects can affect this evolution in a number of ways. However, it can be seen that this 
multifluid set-up does not prevent the development of the classic vortex in the neutral fluid (see Fig. 
\ref{weak_32wide_NI3_v_80.eps}). 

\begin{figure}
\centering
\includegraphics[width=8.4cm]{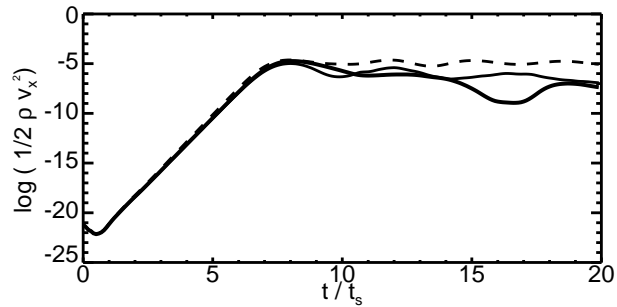}
  \caption{The log of the transverse kinetic energy in the system is plotted against time for the ideal MHD (thin line) and multifluid MHD (thick line) cases. The dashed line represents the hydrodynamic case. There is very little difference between the linear growths for the ideal and multifluid cases.} 
 \label{nonideal_logKEx}
\end{figure}

As described in Paper I, analysis of the transverse kinetic energy in the system allows for study of the growth
rate of the instability. Figure \ref{nonideal_logKEx} plots the growth of the kinetic energy resulting from
the instability in the ideal and multifluid MHD cases, as well as the hydrodynamic case. It can be seen that the linear growth of the instability differs very little with the inclusion of multifluid effects. It will be seen that the instability is developing in a way very different to the ideal MHD case, yet the linear growth rates are found to be within 1\% of each other.

\subsection{Non-linear regime}
\label{subsec:non-linear}

\subsubsection{Diffusion}
\label{subsubsec:diffusion}

\begin{figure}
\centering
\includegraphics[width=8.4cm]{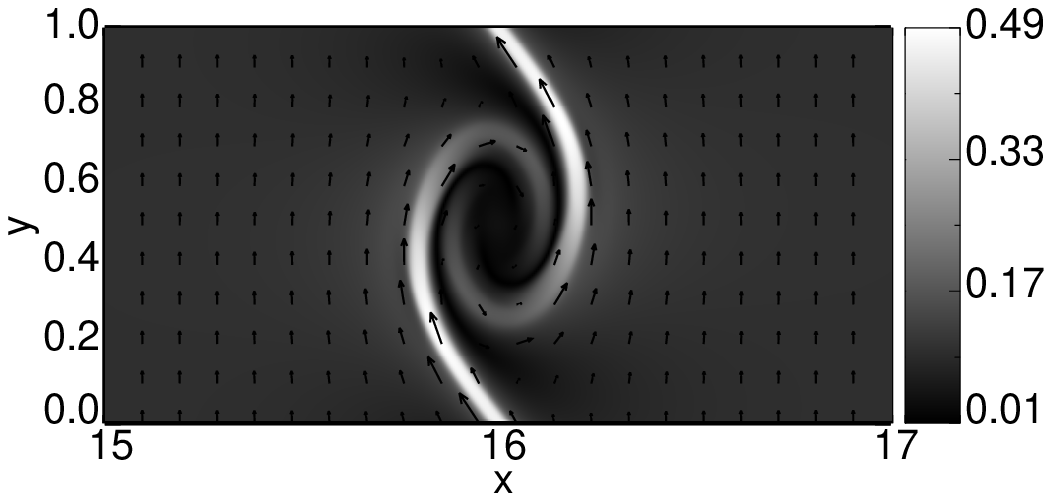}
%
\vspace{0.1cm}
\includegraphics[width=8.4cm]{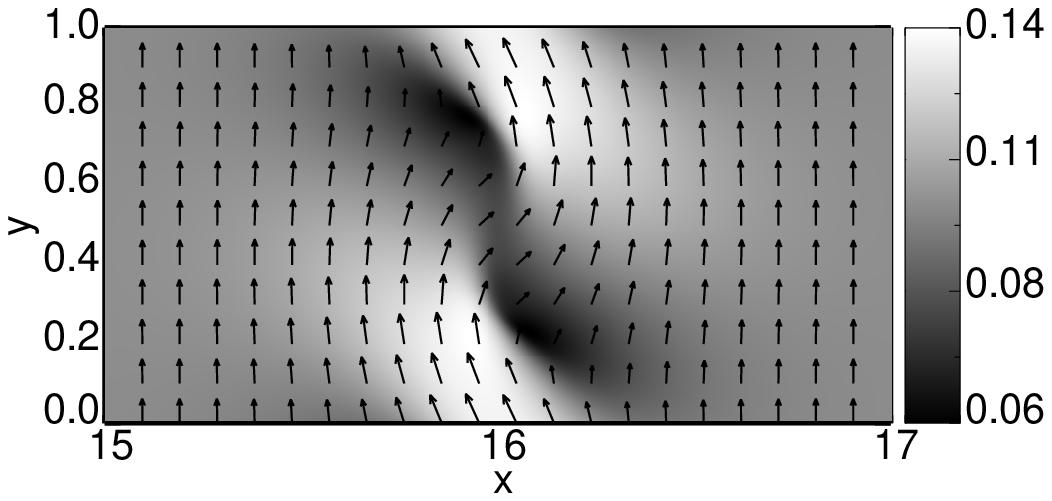}
  \caption{Plot of the magnitude and vector field of the magnetic field for the ideal MHD (upper panel) and multifluid MHD cases (lower panel), at time = $8 \, t_{\rm s}$. In the multifluid MHD case the magnetic field no longer demonstrates a wind-up as seen in the ideal MHD case, as a result of decoupling and diffusion.} 
 \label{nonideal_bfield_vectorplot}
\end{figure}

In the ideal MHD case, the plasma is wound up by the KH instability and the magnetic field experiences a similar winding force as a result of the frozen-in approximation (see Fig. \ref{nonideal_bfield_vectorplot}, upper panel).

It can clearly be seen that the magnetic field undergoes a very different evolution when multifluid effects 
are included (see Fig. \ref{nonideal_bfield_vectorplot}, lower panel).  The inclusion of ambipolar resistivity into 
the system allows for decoupling between the various fluids. This breaks the frozen-in approximation of ideal MHD. 
As a result, the magnetic field is able to diffuse with respect to the bulk fluid. This ambipolar diffusion is the 
source of the altered magnetic field configuration observed. 

\begin{figure}
\centering
\includegraphics[width=8.4cm]{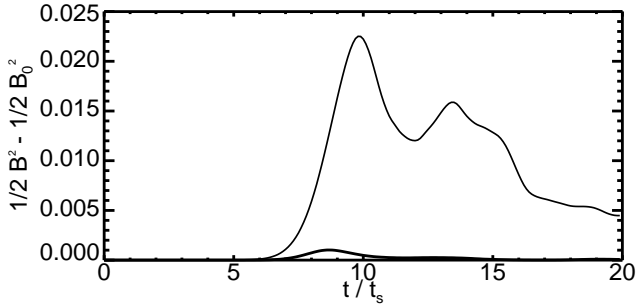}
  \caption{The perturbed magnetic energy in the system is plotted against time in the ideal MHD case (thin line) and multifluid MHD case (thick line). The introduction of multifluid effects significantly dampens the amplification of the magnetic energy.} 
 \label{nonideal_perturbedB}
\end{figure}

The changes in the magnetic field development can be analysed in a more quantitative manner using the plot in figure \ref{nonideal_perturbedB}. The thin line shows the amplification experienced by the magnetic field through the wind-up it undergoes in the ideal MHD case. The thick line shows that this amplification is significantly reduced in the presence of ambipolar diffusion. Similar results were observed in Paper I in the ambipolar-dominated simulation. As this was not observed in the Hall-dominated simulations in Paper I, we can deduce that this is solely as a result of the ambipolar resistivity. 

\subsubsection{Charged fluids}
\label{subsubsec:charged-fluids}

\begin{figure}
\centering
\includegraphics[width=8.4cm]{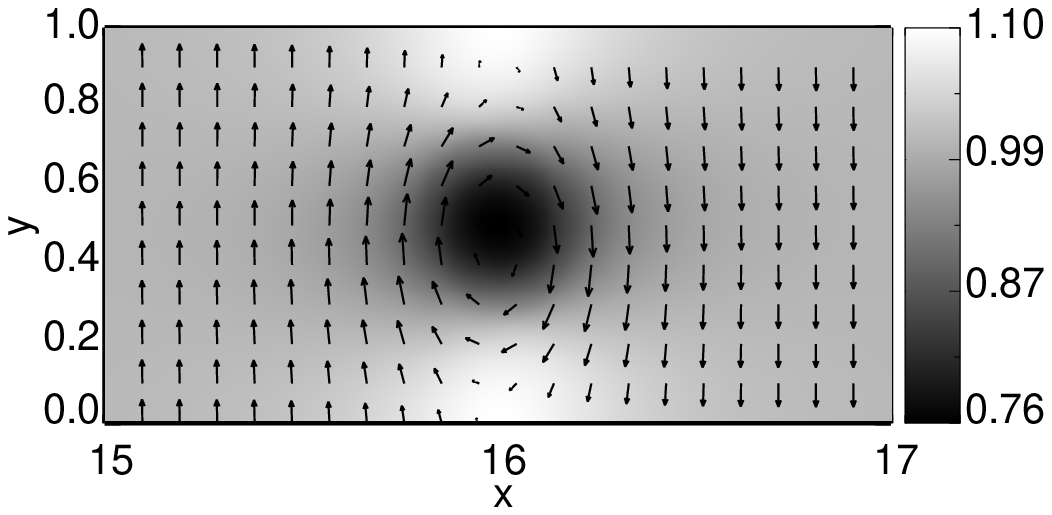}
%
%
\vspace{0.1cm}
\includegraphics[width=8.4cm]{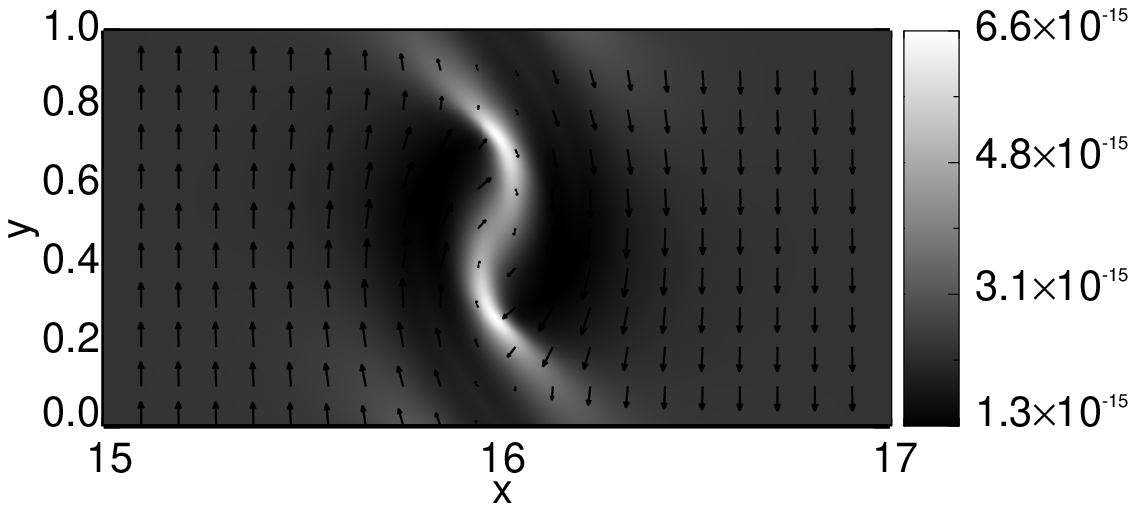}
%
%
\vspace{0.1cm}
\includegraphics[width=8.4cm]{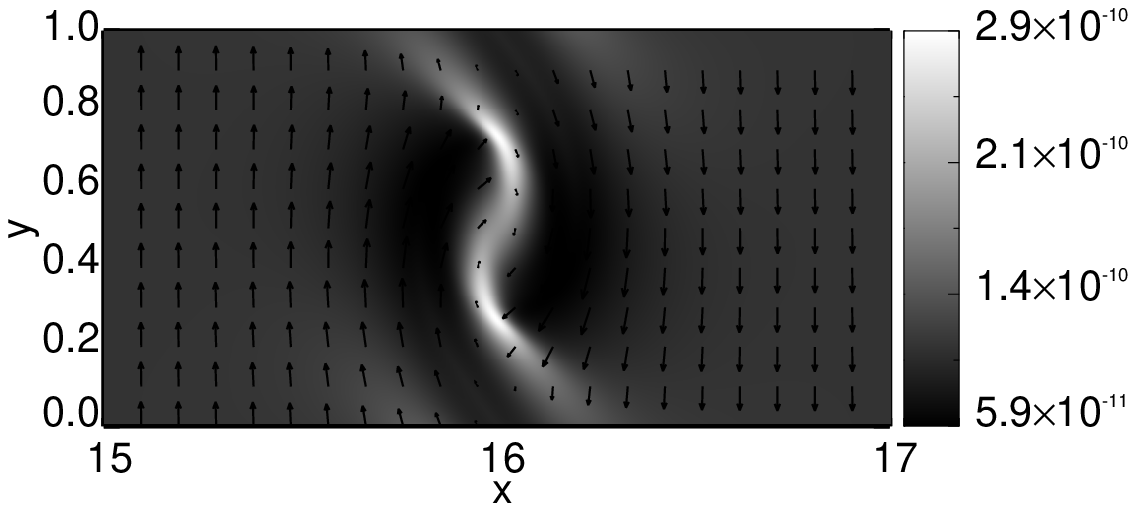}
%
\vspace{0.1cm}
\includegraphics[width=8.4cm]{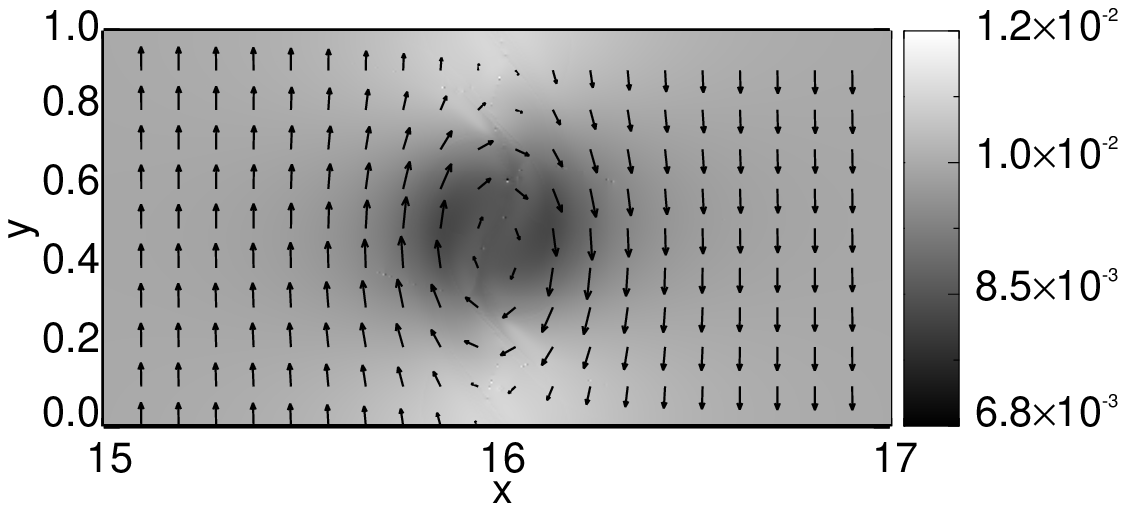}
  \caption{Plot of the distribution of the mass density, as well as the
	  velocity vector field, of the bulk fluid (top panel), electron fluid (second panel), ion fluid (third panel) and the dust grain fluid (bottom panel) in the multifluid case at time $=8\,t_{\rm s}$.} 
 \label{results_1KH_density_den5}
\end{figure}

We know that the introduction of ambipolar resistivity has allowed for decoupling of the magnetic field from
the neutral fluid. We now examine the behaviour of the charged fluids themselves.  The exact behaviour of each
charged fluid can be understood by examining its density profile and velocity field during the development of the instability. The state of each of the four fluids in the system has been plotted in figure \ref{results_1KH_density_den5} at the time of saturation of the instability.

It can be seen that the mass density of the dust grain fluid closely reflects that of the bulk fluid,
signifying a strong coupling between the two.  This would be expected due to its relatively low Hall
parameter (see equation \ref{eqn:dust_hall}).  On the other hand, the ion and electron fluids more closely
reflect the configuration of the magnetic field, implying that they are still strongly coupled to the
magnetic field lines, as expected by their high Hall parameter (equations \ref{eqn:ion_hall} and
\ref{eqn:electron_hall}). 
The decoupling of the ion and electron fluids from the neutral fluid is the source of the ambipolar diffusion in the 
system. The magnetic field is tied to the neutral fluid only through the coupling of the charged fluids with
the neutrals, so a low collisional coupling between the charged fluids and the neutrals allows for the magnetic field to diffuse relative to the bulk fluid. 

\begin{figure}
\centering
\includegraphics[width=8.4cm]{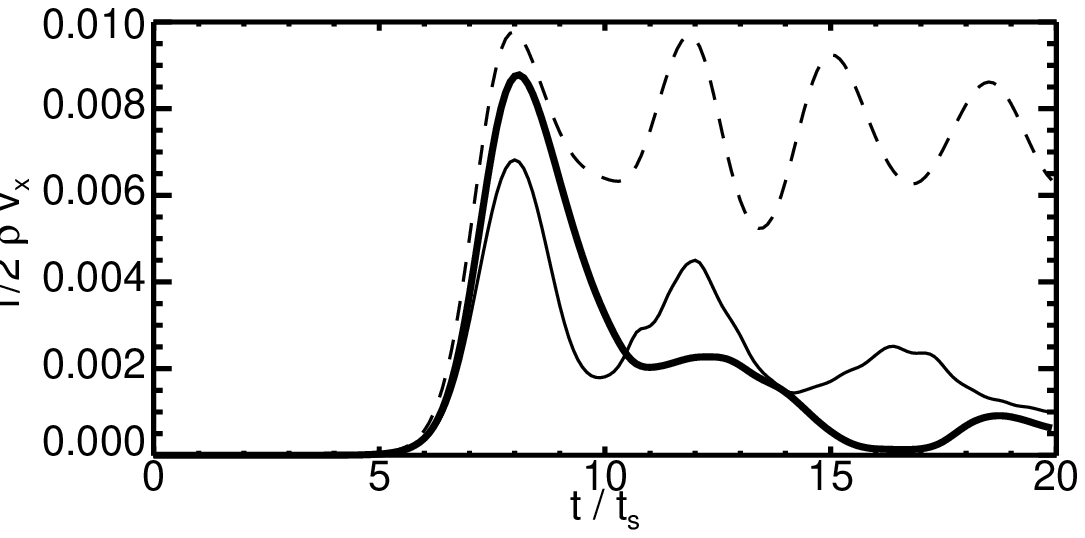}
%
%
%
\vspace{0.1cm}
\includegraphics[width=8.4cm]{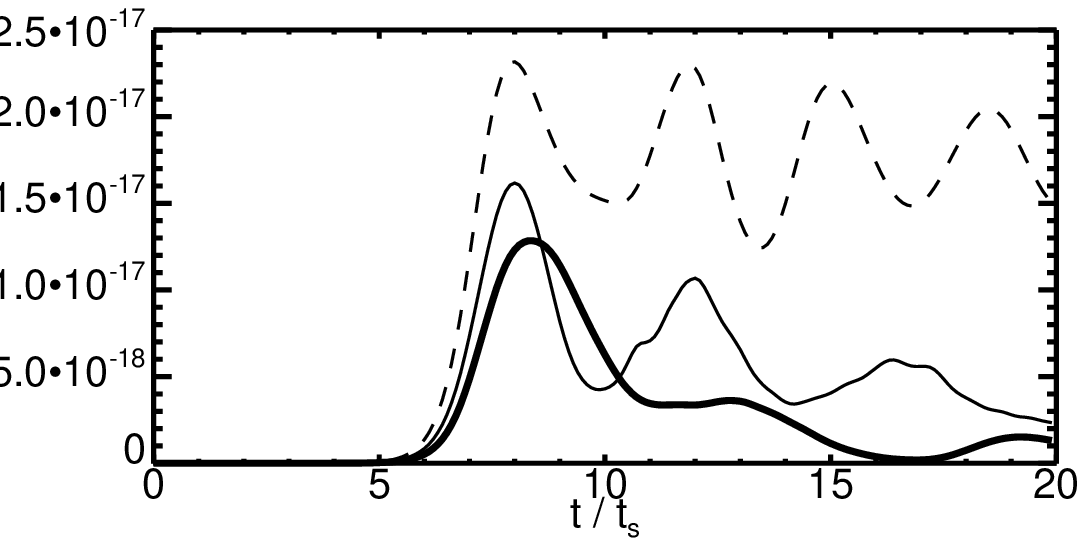}
%
\vspace{0.1cm}
\includegraphics[width=8.4cm]{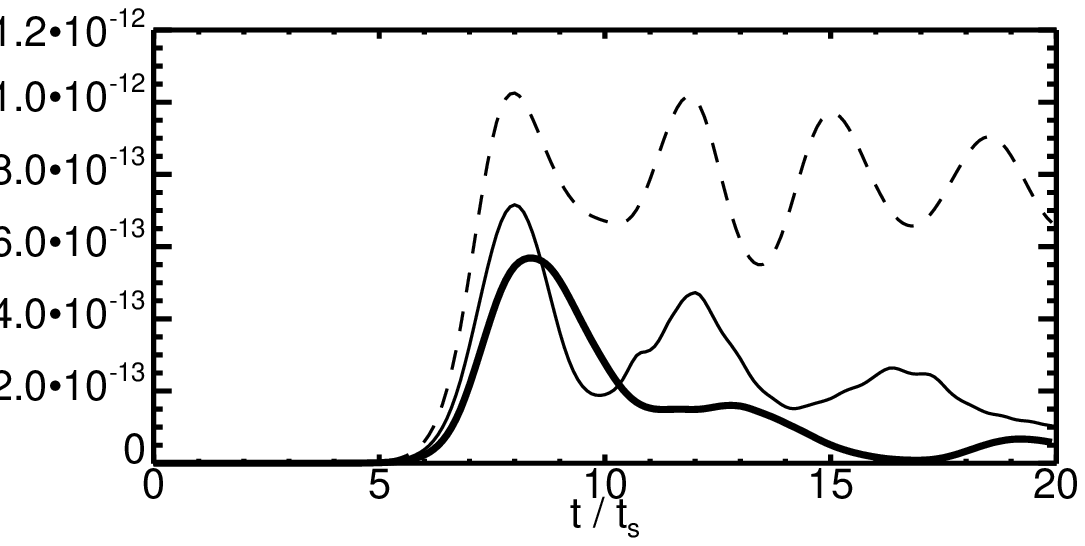}
%
\vspace{0.1cm}
\includegraphics[width=8.4cm]{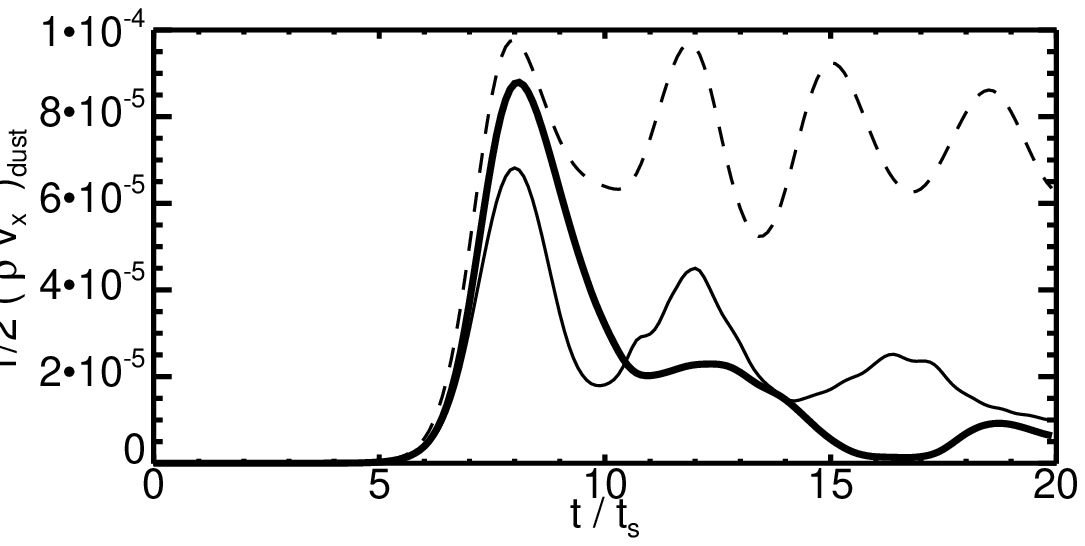}
  \caption{The transverse kinetic energy is plotted against time for the ideal MHD (thin line), multifluid MHD (thick line) and hydrodynamic cases (dashed line), for the entire system (top panel), electron fluid (second panel), ion fluid (third panel) and dust grain fluid (bottom panel).} 
 \label{nonideal_KEx4}
\end{figure}

The various dynamics discussed above are confirmed in figure \ref{nonideal_KEx4}. These plots of the
transverse kinetic energy for each of the four fluids clearly demonstrate the behaviour of each. We can see
that the bulk fluid undergoes further wind-up in the multifluid MHD case than in the ideal MHD case (see the
top panel of figure \ref{nonideal_KEx4}). This is
due to two distinct phenomena (see Paper I for more details). In general, the system is prevented from as
strong a wind-up as seen in the hydrodynamic case by the presence of a magnetic field. In multifluid MHD,
there are two effects at work that limit the effectiveness of the magnetic field in suppressing this wind-up.
Firstly, the introduction of even a small amount of ambipolar diffusion causes the magnetic field to
experience a significant reduction in its amplification. The resulting weaker magnetic field allows the bulk
fluid to undergo a stronger wind-up. Secondly, with higher levels of ambipolar diffusion being introduced into
the system, the bulk fluid becomes further decoupled from the magnetic field, further reducing its effectiveness in opposing the wind-up.

On the other hand both the electron and ion fluids experience a decoupling from the neutral fluid. As they are
still well tied to the magnetic field, and the magnetic field no longer winds up in a manner similar to the
bulk fluid, the charged fluids undergo less wind-up than in the ideal MHD case. This is demonstrated by the
plots of the transverse kinetic energy of these fluids (figure \ref{nonideal_KEx4}, second and third panels),
and the lower maxima reached. Finally, the plot representing the dust grain fluid (figure
\ref{nonideal_KEx4}, bottom panel) shows similar behaviour to that observed in the bulk fluid. This confirms that the dust grains remain well-coupled to the neutral fluid, and are less affected by the magnetic field than the other charged fluids.
 

\subsubsection{Combination of multifluid effects}
\label{subsubsec:combo}

\begin{figure}
\centering
\includegraphics[width=8.4cm]{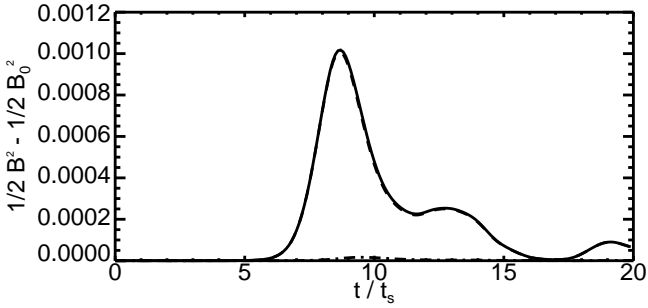}
  \caption{The total perturbed magnetic energy (solid line) in the system is plotted against time in the multifluid MHD case. This is almost identical to the magnetic energy in the $xy$-plane (dashed line), while there is very little growth in the magnetic energy in the $z$-direction (dot-dashed line).} 
 \label{hall_ideal_bxyz2}
\end{figure}

The results detailed above can be attributed to the introduction of ambipolar diffusion into the system. However, there is a significant amount of Hall resistivity included in the set-up as well. While the system remains ambipolar-dominated, results from Paper I indicate that there should be observable consequences of including this moderate amount of Hall resistivity. Most notably, a small but significant re-orientation of the magnetic field is expected, from the $xy$-plane into the $z$-direction. Plotting the evolution of the magnetic energies in each of the three directions however, shows that there is, in fact, no significant growth in the $z$-direction (see figure \ref{hall_ideal_bxyz2}). This is in distinct opposition with the results from Paper I, in which significant growth of the magnetic energy in the $z$-direction was observed for the same level of Hall resistivity. 

It is important to recall that the strength of the Hall effect depends on the current in the system. The current, in turn, depends on the charge densities of the three charged fluids. The charge density of the dust grain fluid is $\alpha_3 \rho_3 \approx -8 \times 10^{-18} \,\mbox{statC\,cm}^{-3}$. However, for the ion and electron fluids, the charge densities are much higher, approximately $\pm 5 \times 10^{-15} \, \mbox{statC\,cm}^{-3}$. Therefore the current in the system, $\mathbf{J} = \sum_{i} \alpha_i \rho_i \mathbf{v}_i$, is primarily due to the velocity difference between the ion and electron fluids. 

Therefore, while the decoupling of the dust grain fluid from the magnetic field provides a high Hall
resistivity, the strength of the Hall effect is in fact also critically dependent on the dynamics of the
ion fluid relative to the electron fluid. As the ion fluid Hall parameter is much greater than 1 it will be somewhat 
decoupled from the neutral fluid, instead being more strongly tied to the magnetic field. This means that the 
relative velocity between the ion and electron fluid will be rather small. As a result, the majority of
current in the system remains {\em parallel} to the magnetic field, and the strength of the Hall effect, which
is proportional to $\mathbf{J} \times \mathbf{B}$, is very small. This accounts for the growth of the magnetic
field in the $z$-direction being much less than would naively be expected. Put simply, the introduction of
ambipolar resistivity changes which fluids are coupled to which and, in doing so, inhibits the Hall effect.

On a side note, this can also occur in a system with only two charged fluids. In this case, it is typically
the ion fluid that is the larger contributor to both the Hall and ambipolar conductivity. The decoupling of
the ion fluid from the magnetic field leads to Hall resistivity, as any velocity difference between the ion
and electron fluid can give rise to a current with a component perpendicular to the magnetic field. However,
the ambipolar resistivity arises due to a low collision rate between the ion fluid and the neutral fluid.
As a result, the ions do not, in fact, behave in the same way as the neutrals.  Their coupling with the
magnetic field, though weak, is sufficient to ensure only a minimal relative velocity between the ions and 
electrons. In this way, the impact of the Hall effect is similarly reduced.

\begin{figure}
\centering
\includegraphics[width=8.4cm]{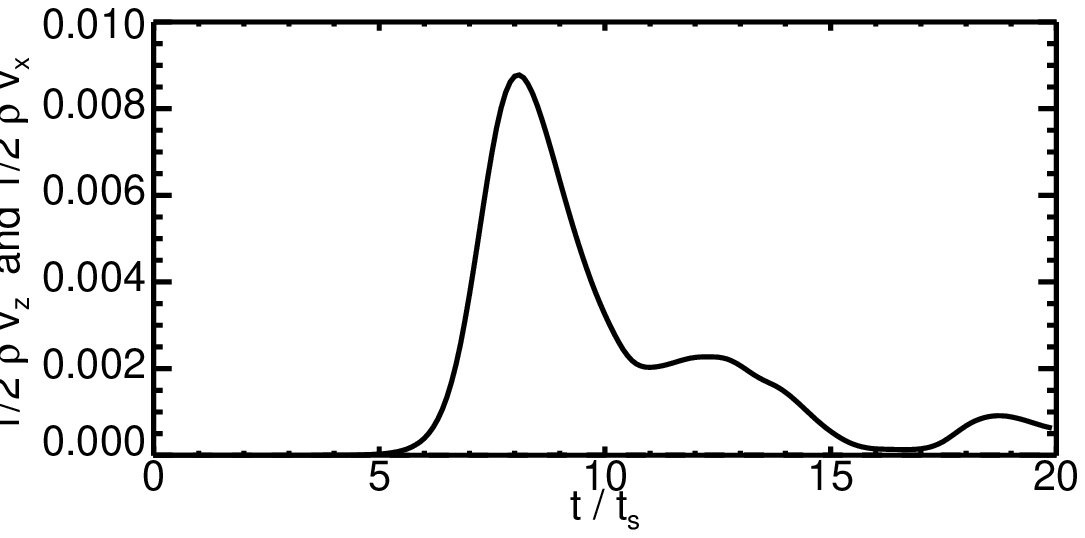}
%
\vspace{0.1cm}
\includegraphics[width=8.4cm]{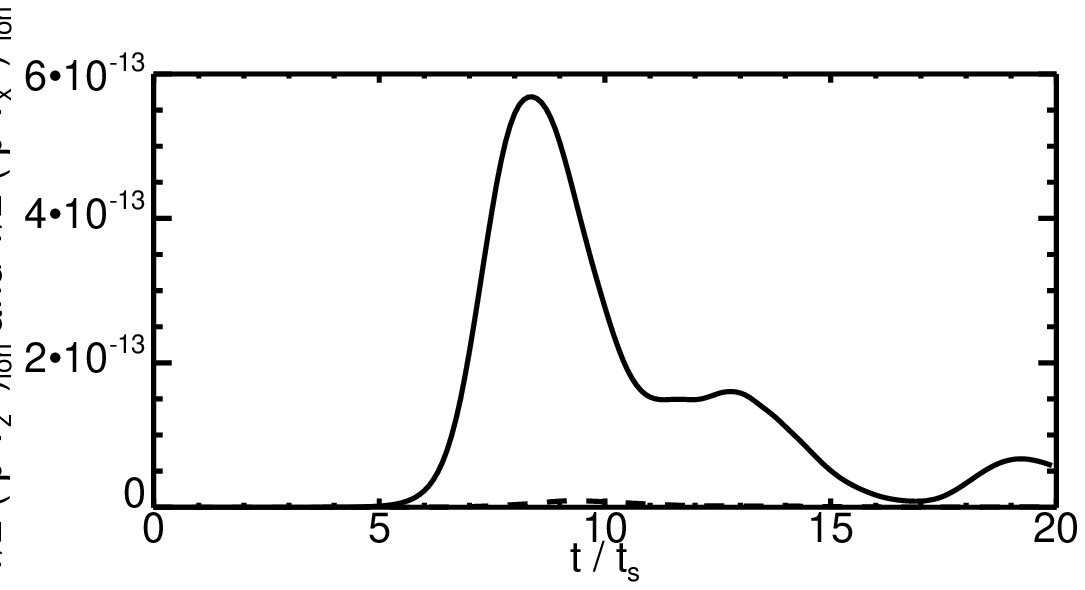}
  \caption{The growth of the kinetic energy of the entire system (upper panel) and the ion fluid (lower panel) is plotted against time in both the $x$-direction (solid line) and $z$-direction (dashed line). The growth in the $z$-direction for the bulk flow is clearly negligible, while the growth in the $z$-direction for the ion fluid is very small, but non-negligible.}
 \label{hall_ideal_kx_perturbB+kz_bz2}
\end{figure}

In a Hall-dominated flow, we would expect to see not only a growth in magnetic field strength in the $z$-direction, but also a resulting growth of kinetic energy in the same direction, as the plasma is influenced to travel out of the $xy$-plane. With even moderate Hall resistivity, this kinetic energy in the $z$-direction can become comparable to that in the $x$-direction (see Paper I). However, in this multifluid case, the magnetic field experiences only a minimal re-orientation into the $z$-direction, and therefore only the fluids that are tightly coupled to the magnetic field will experience any noticeable dynamics in this direction. Figure \ref{hall_ideal_kx_perturbB+kz_bz2} shows that the bulk flow demonstrates negligible growth of kinetic energy in the $z$-direction (upper panel), while the ion fluid, being more closely tied to the magnetic field, demonstrates a very small, but non-negligible, growth in this direction (lower panel).

\subsection{Subsequent behaviour}
\label{subsec:subsequent}

\begin{figure}
\centering
\includegraphics[width=8.4cm]{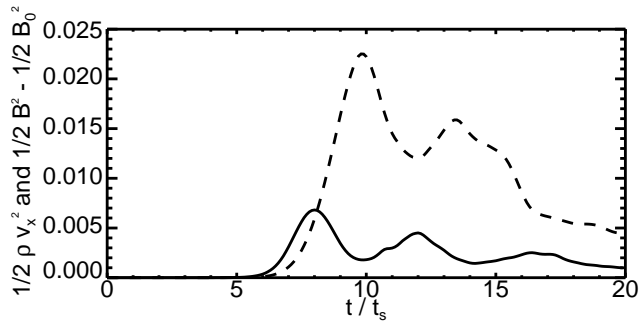}
  \caption{The growth of the perturbed magnetic energy (dashed line) is plotted against time in the ideal case. Also plotted is the transverse kinetic energy (solid line) of the ideal case. It can be seen that, after the initial saturation point, the growth of magnetic energy is linked to the decay of kinetic energy, and vice-versa.} 
 \label{results2_ideal_magE_KEx}
\end{figure}

\begin{figure}
\centering
\includegraphics[width=8.4cm]{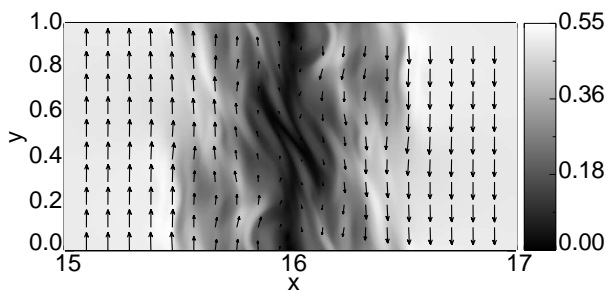}
  \caption{Plot of the magnitude and vector field of the velocity field for the ideal MHD case, at time = $20 \, t_{\rm s}$. The system has become somewhat disordered.} 
 \label{results_2KH_KEx_vmag_later}
\end{figure}

The KH instability is seen to undergo a very different evolution in the presence of multifluid effects. In the ideal case, the initial wind-up of the velocity field has the effect of also winding up the magnetic field, due to strong coupling between the two. As the KH vortex is caused to stretch and expand by the amplified magnetic field, it reaches the periodic $y$-boundaries of the simulated grid. As the neighbouring vortices merge, numerical viscosity allows for magnetic reconnection, which results in the creation of magnetic islands and secondary vortices. The generation and decay of these vortices results in further periods of growth in the transverse kinetic energy in the system corresponding to periods of decreasing magnetic energy and vice-versa. This is clearly demonstrated in figure \ref{results2_ideal_magE_KEx}, where the peaks in each energy are seen to correspond to the troughs in the other. This behaviour continues in the ideal MHD system until the system has reached a somewhat disordered state (see figure \ref{results_2KH_KEx_vmag_later}).

\begin{figure}
\centering
\includegraphics[width=8.4cm]{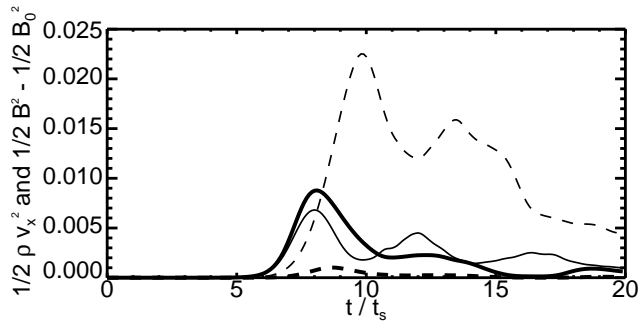}
  \caption{The evolution of the transverse kinetic energy (solid lines) and magnetic energy (dashed lines) for the ideal MHD case (thin lines) and multifluid case (thick lines). It can be seen that, after the initial saturation point,the magnetic field has experienced little growth, and is unable to inject energy back into the system.} 
 \label{perturbB_KEx_ideal_ambi}
\end{figure}

\begin{figure}
\centering
\includegraphics[width=8.4cm]{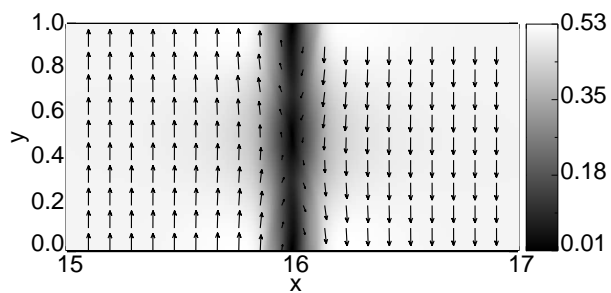}
  \caption{Plot of the magnitude and vector field of the velocity field for the high ambipolar case, at time =
	  $20 \, t_{\rm s}$. The system has returned to almost laminar flow.} 
 \label{results_1KH_KEx_vmag_later}
\end{figure}

In the multifluid case, on the other hand, the subsequent evolution of the KH instability is quite different.
The high ambipolar resistivity causes the magnetic field to experience very little wind-up, through decoupling
and diffusion. As a result, the KH vortex undergoes little or no stretching or expansion as a consequence of
the magnetic field. Following the saturation of the instability, the magnetic field is seen to steadily return
to its original configuration, and neither the velocity field nor the magnetic field undergo a second period
of growth (see figure \ref{perturbB_KEx_ideal_ambi}). Instead, the magnetic field eventually stabilises the
velocity field, and the plasma returns almost to its original state, with simple laminar flow on either side
of a much wider shear layer (see figure \ref{results_1KH_KEx_vmag_later}). This wider shear layer is no longer 
conducive to the growth of the KH instability.

\begin{figure}
\centering
\includegraphics[width=8.4cm]{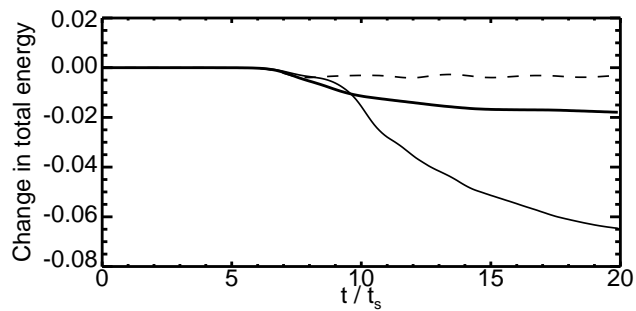}
  \caption{The total energy in the system of both the ideal MHD (thin line) and multifluid MHD (thick line) systems are plotted against time. While the ideal case continues to lose energy throughout the simulation through disordered decay, the levelling off of the energy in the multifluid case indicates that it has reached a stable state.} 
 \label{nonidealHD_KEy}
\end{figure}

As the vortex is broken down and the system returns to a stable state, the energy consumed by the instability
flattens out. The system has lost some energy during the initial growth of the instability due to ambipolar
diffusion removing some magnetic energy, but this levels off at later times. This is a deviation from the
ideal MHD case, in which the system continues to lose energy as the KH vortex is broken down through
reconnection into disordered decay. Figure \ref{nonidealHD_KEy} plots the total energies of the ideal and
multifluid MHD cases, as well as the hydrodynamic case. In the hydrodynamic case, the KH vortex remains
indefinitely, as there is no magnetic field to lead to its decay. For these reasons, the total energies in the multifluid and hydrodynamic cases are seen to level off, while the ideal MHD system continues to lose energy through decay.

\section{Conclusions}

A study of the KH instability in a molecular cloud plasma is carried out. We restrict our
attention to the case of a single, transonic, super-Alfv\'enic shear layer and, as such, the body modes of
	the KH instability are not examined.  The KH instability,
in particular the growth of its surface modes, is severely reduced for high Mach number flows,
while sub-Alfv\'enic flows tend to be stabilised by the presence of the magnetic field.  Our choice
of an Alfv\'en number of 10 for the system is realistic in terms of that present in a molecular
cloud and means that the growth of the instability, in the ideal MHD approximation where there is
nearly no resistivity, will roll up the magnetic field eventually leading to a situation in which 
tearing mode instabilities can reorganise the topology of the magnetic field \citep{jones97,
baty03}.  We utilise periodic boundary conditions in the longitudinal direction and, while a
very usual approach, it should be remembered that the so-called ``inverse cascade'' observed by
\citet{baty03} will not occur in this system.

The combined effects of ambipolar diffusion and the 
Hall effect present in such a plasma are examined with reference to an ideal MHD system. As molecular 
clouds are dominated by ambipolar diffusion, it is expected that the results will closely reflect those 
of the ambipolar-dominated simulation in Paper I and this is indeed observed.  However, one's naive 
expectation would be that the Hall effect should also be observed through twisting of the magnetic field 
lines. In the initial linear regime, the growth rates of the instability remain unchanged by multifluid 
effects. As the instability develops, ambipolar diffusion leads to less amplification of the magnetic 
energy in the system and therefore a stronger wind-up of the neutral fluid. This is very similar to the 
results observed in the ambipolar-only system studied in Paper I.  Subsequent behaviour of the 
instability, including the breakdown of the KH vortex and its return to a stable state and laminar flow, 
also closely reflects that observed in the ambipolar-only system.

A noteworthy result of this study is the lack of the expected impact on the system due to the inclusion 
of the Hall effect. This appears to be due to the presence of ambipolar diffusion: i.e.\ ambipolar 
diffusion {\em inhibits} the impact of the Hall effect.  The set-up presented here includes
three charged fluids (ions, electrons and dust grains) as well as the neutral fluid.
However, this same result can be observed in a system with two charged fluids (ions and electrons) as 
well as the neutral fluid. In the case of two charged fluids, the Hall parameter of the ions
will be much lower than that of the electrons, thereby introducing Hall resistivity. Ambipolar 
resistivity then arises when the ion fluid also has a low collision rate with, or more properly momentum 
transfer to, the neutral fluid. As the ion fluid is subjected to fewer collisions with the neutrals, it 
is the magnetic field that plays the dominant role in determining its dynamics, despite the relatively 
weak coupling. This minimises the impact of the Hall effect in the system. In the case of three charged 
fluids, Hall resistivity arises in the system from the dust grain fluid being weakly coupled to the 
magnetic field.  Ambipolar diffusion will arise when the collision rate of the ion fluid with the 
neutrals is low. However, when the dust grain fluid has a much smaller charge density than that of the 
ion and electron fluids, the effective current is due mainly to the relative velocities of the ion and 
electron fluids and not the dust grain fluid. As both the electron and ion fluids remain tightly coupled 
to the magnetic field, their relative velocity will continue to be in the direction of the magnetic 
field lines, resulting again in a suppression of the Hall effect in the system.

We conclude that for the parameters observed in ambipolar-dominated molecular clouds, the evolution of 
the KH instability is indeed dominated by the presence of ambipolar diffusion. However, the system does 
not demonstrate the expected twisting of the magnetic field lines which would arise from the inclusion 
of the Hall effect as naively expected.  This suppression of the Hall effect by ambipolar diffusion is 
given an original and detailed explanation based on basic physics principles and for plasmas involving a 
variety of charged fluids.

Finally, we note that it would be fascinating, though computationally very expensive, to perform 
very large scale simulations in which the inverse cascade seen by \cite{baty03} can transfer energy from 
length scales on which multifluid effects are important right up to length scales on which ideal MHD is 
an appropriate approximation in order to study the nature of this cascade in detail.  For the work
presented here, however, our chosen length scales of $\sim 0.2$\,pc are appropriate for the study of the
effect of this instability on the bowshocks caused by young stellar outflows.

\section*{Acknowledgements}
The research of A.C.J. has been part supported by the CosmoGrid project funded 
under the Programme for Research in Third Level Institutions (PRTLI) 
administered by the Irish Higher Education Authority under the National 
Development Plan and with partial support from the European Regional 
Development Fund.

The authors wish to acknowledge the SFI/HEA Irish Centre for High-End
Computing (ICHEC) for the provision of computational facilities and
support.

\label{lastpage}

\end{document}